\documentclass[]{pasj01}
\draft

\begin{document} 

\title{A Large Convective-Core Overshoot in \textit{Kepler} Target KIC 11081729}

\author{Wuming \textsc{Yang}\altaffilmark{1}%
}
\altaffiltext{1}{Department of Astronomy, Beijing Normal University, Beijing 100875, China}
\email{yangwuming@bnu.edu.cn, yangwuming@ynao.ac.cn}


\maketitle

\begin{abstract}
The frequency ratios $r_{01}$ and $r_{10}$ of KIC 11081729 decrease firstly
and then increase with the increase in frequency. For different spectroscopic
constraints, all models with overshooting parameter $\delta_{\mathrm{ov}}$ 
less than 1.7 can not reproduce the distributions of the ratios. 
However, the distributions of the ratios can be directly reproduced 
by models with $\delta_{\mathrm{ov}}$ in the range of about $1.7-1.8$. 
The estimations of mass and age of the star can be affected by spectroscopic 
results, but the determination of the $\delta_{\mathrm{ov}}$ is not dependent
on the spectroscopic results. A large overshooting of convective core may 
exist in KIC 11081729. The characteristics of $r_{01}$ and $r_{10}$ 
of KIC 11081729 may result from the effects of the large overshooting 
of convective core. The distributions of $r_{01}$ and $r_{10}$ of different
stars with a convective core can be reproduced by the function $B(\nu_{n,1})$. 
If the value of the critical frequency $\nu_{0}$ is larger than the value 
of frequency of maximum oscillation power $\nu_{max}$, a star may have 
a small convective core and $\delta_{\rm ov}$. But if the value of 
$\nu_{0}$ is less than that of $\nu_{max}$, the star may have a large 
convective core and $\delta_{\mathrm{ov}}$. 
The function aids in determining the presence of convective core and the size 
of the convective core including overshooting region from observed frequencies.
The determination is not dependent on the calculation of stellar models. 
\end{abstract}

\section{Introduction}

Asteroseismology has proved to be a powerful tool for determining
the fundamental parameters of stars, diagnosing the internal structure
of stars, and probing physical processes in stellar interiors
by comparing observed oscillation characteristics with those calculated
from theoretical models \citep{mazu06, cunh07, chri10, yang12, silv13, liu14,
guen14, chap14, metc14, tian14}. 

However, the stellar model found solely by matching individual 
frequencies of oscillation does not always properly reproduce 
other characteristics of oscillations \citep{dehe10, silv13, liu14}, 
such as the ratios of small-to-large separations, $r_{10}$ and $r_{01}$,
which are sensitive to conditions in stellar core \citep{roxb03}.
The small separations are defined as \citep{roxb03}
\begin{equation}
d_{10}(n)\equiv-\frac{1}{2}(-\nu_{n,0}+2\nu_{n,1}-\nu_{n+1,0})
\label{d10}
\end{equation}
and
\begin{equation}
d_{01}(n)\equiv\frac{1}{2}(-\nu_{n,1}+2\nu_{n,0}-\nu_{n-1,1}).
\label{d01}
\end{equation}
In calculation, equations (\ref{d10}) and (\ref{d01}) are
rewritten as the smoother five-point separations.

Stars with a mass larger than $1.1$ $M_\odot$ may have a convective 
core during their main sequence (MS). The overshooting of the 
convective core extends the region of material mixing 
by a distance $\delta_{\rm ov}$ $H_{p}$ above the top of the 
convective core that is determined by Schwarzchild criterion,
where $H_{p}$ is the local pressure scale-height and $\delta_{\rm ov}$
is a free parameter. The overshooting brings more H-rich material
into the core, prolongs the lifetime of the burning
of core hydrogen, and changes the internal structure of
stars and the global characteristics of
the following giant stages \citep{schr97, yang12}.
The determination of the global parameters of stars 
by asteroseismology or other studies based on stellar 
evolution can be directly affected by the overshooting 
\citep{mazu06, yang12}. 

However, for stars with a mass of around $1.1$ $M_\odot$,
there may or may not exist a convective core in their interior,
depending on the input physics used in the computation of
their evolutions \citep{chri10}. Up to now, there is no a direct
method to determine the presence and the size of the convective core 
and the extension of overshooting from observed data. Therefore,
finding a method of determining the presence and the 
size of the convective core is important for understanding 
the structure and evolution of stars.

\citet{dehe10} argued that the presence or absence of a convective core
in stars can be indicated by the small separations and that the star 
HD 203608 with $M\simeq0.94$ $M_\odot$ and $t\simeq6.7$ Gyr in fact
has a convective core. \citet{deme10} studied the oscillations 
of $\alpha$ Centauri A ($1.105\pm0.007$ $M_\odot$) and concluded 
that the $d_{01}$ allows them to set an upper limit to the amount 
of convective-core overshooting and that the model of $\alpha$ 
Centauri A with a radiative core reproduces the observed $r_{01}$ 
significantly better than the model with a convective core.
Moreover, \citet{silv13} tried to detect the convective core of 
KIC 6106415 and KIC 12009504. They obtained that the mass of KIC 6106415
is $1.11\pm0.05$ $M_\odot$ and that of KIC 12009504 is $1.15\pm0.04$ $M_\odot$,
and concluded that a convective core and core overshooting exist in KIC 12009504,
but could not determine whether a convective core exists in KIC 6106415.
\citet{tian14} studied the oscillations of KIC 6225718 and concluded
that either a small convective core or no convective core exists 
in the star.

For MS stars, the ratios generally decrease with frequency.
\citet{liu14} argued that the ratios affected by the overshooting
of convection core could exhibit an increase behavior.
By using this seismic tool, the value of overshooting parameter
$\delta_{\rm ov}$ is restricted between 0.4 and 0.8 for HD 49933 \citep{liu14}
and between 1.2 and 1.6 for KIC 2837475 \citep{yang15}. 
Moreover, \citet{guen14} found that the value of $\delta_{\rm ov}$ is between 
0.9 and 1.5 for Procyon, which is supported by the work of \citet{bond15}.
The values of $\delta_{\rm ov}$ are larger than the value of $0.1-0.3$ 
estimated by comparing the theoretical and observational
color-magnitude diagram of star clusters \citep{prat74, dema94} 
and characteristics of eclipsing binary stars \citep{schr97}.

However, the presence of the large overshooting is supported
by the convective theory of stars and numerical simulation.
For example, \citet{xiong85} shows that if
the overshooting distance is defined as the distance up to 
which mixing of matter extends, the overshooting distance of
the convective core can reach $1.4c_{1}$ $H_{p}$ in his 
convective theory, where the value of the $c_{1}$ is 
between $1/3$ and $1$; the numerical simulation of \citet{tian09}
shows that the penetration distance of downward overshooting of
convection of giant stars can reach $1-2$ $H_{p}$. Moreover, 
\citet{vall16} show that the calibration of the convective
core overshooting of double-lined eclipsing binaries with the mass
in the range of $1.1-1.6$ $M_\odot$ is poorly reliable. They suggested
that asteroseismic data may be required for the calibration 
of $\delta_{\rm ov}$.

KIC 11081729 is an F5 star \citep{wrig03}. The frequencies 
of p-modes of KIC 11081729 have been extracted by \citet{appo12}.
Combing asteroseismical and non-asteroseismical data,
several authors \citep{hube14, metc14} have investigated
the fundamental parameters of KIC 11081729. The mass determined
by \citet{hube14} is $1.36^{+0.04}_{-0.06}$ $M_\odot$ for KIC 11081729,
but that determined by \citet{metc14} is $1.26\pm0.03$ $M_\odot$. 
The observed ratios $r_{10}$ and $r_{01}$ of KIC 11081729
decrease firstly and then increase with the increase in frequencies.
The characteristics may derive from the effects of overshooting
of convective core. If these characteristics can be directly 
reproduced by stellar models, which may aid in understanding
the origin of the characteristics and confirming the presence
of the large overshooting of convective core.

In this work, we focus mainly on examining whether the observed 
characteristics of KIC 11081729 can be directly reproduced by 
stellar models and can be explained by overshooting of convective core.
In order to find the best model of KIC 11081729, the chi squared method
was used. First, an approximate set of models was found out with the 
constraints of luminosity, $T_{\rm eff}$, and [Fe/H]. Then around this set 
of solutions, we sought for best models that match both non-seismic
constraints and the individual frequencies extracted by \citet{appo12}, 
and then we compared the observed $r_{10}$ and $r_{01}$ with those 
calculated from models. And finally, a function $B(\nu_{n,1})$ 
was deduced and tested. In Section 2, stellar models of KIC 11081729
are introduced. In Section 3, the function $B(\nu_{n,1})$ is deduced, 
and in Section 4, the results are summarized and discussed.


\section{STELLAR MODELS}
\subsection{Input Physics}
\begin{figure}
\centering
\includegraphics[scale=0.34, angle=-90]{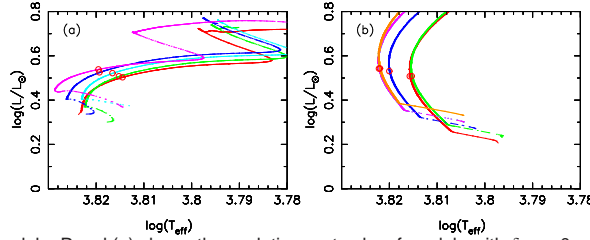}
\caption{Hertzsprung-Russell diagram of models. Panel (a)
shows the evolutionary tracks of models with $\delta_{\rm ov}$ = 0 and 0.2
with the spectroscopic constraints of Bruntt et al. (2012),
while panel (b) depicts those of models with $\delta_{\rm ov}$ = 1.8.
The circles indicate the most likely models for KIC 11081729.}
\label{pghr}
\end{figure}

In order to obtain the best model of KIC 11081729, we construct a grid of 
evolutionary models using the Yale Rotation Evolution Code (YREC)
\citep{pins89, yang07}. For the microphysics, the OPAL equation-of-state 
table EOS2005 \citep{roge02}, OPAL opacity table GN93 \citep{igle96}, and 
the low-temperature opacity tables of \citet{alex94} are used. 
The standard mixing-length theory is adopted to treat convection.
The value of the mixing-length parameter $\alpha$ calibrated to the Sun
is 1.74 for the YREC. But it is a free parameter in this work. 
The full mixing of material is assumed in overshooting region of models.
The diffusion and settling of both helium and heavy elements are taken 
into account in models with a mass less than 1.30 $M_\odot$ by using 
the diffusion coefficients of \citet{thou94}. In the model calculation, 
the initial helium mass fraction is fixed at $0.248$ and $0.295$. 
The value of $0.248$ is the standard big bang nucleosynthesis value \citep{sper07}. 
The values of input parameters, mass ($M$), $\alpha$ and $\delta_{\rm ov}$, 
and heavy-element abundance ($Z$) of Zero-Age MS models are listed in Table \ref{tab1}.
The models are constructed from ZAMS to the end of MS or the subgiant stage.
Some of evolutionary tracks of the models are shown in Figure \ref{pghr}
as an example.

\begin{table*}
\begin{center}
\caption{Input parameters for model tracks.
The symbol $\delta$ indicates the resolution of the parameters. 
\label{tab1}}
\begin{tabular}{ccccc}
  \hline\noalign{\smallskip}
  \hline\noalign{\smallskip}
         Variable    & Minimum  & Maximum & $\delta$  \\
   \hline\noalign{\smallskip}
         M/M$_\odot$ & 1.01     & 1.50   & $\leq$0.02    \\
        $\alpha$     & 1.65     & 2.15 &  0.1     \\
        $\delta_{\rm ov}$ &  0.0     & 1.8$^{a}$ &  0.2     \\
        $Z_{i}$      & 0.010    & 0.040 & 0.002   \\
   \hline\noalign{\smallskip}
  \noalign{\smallskip}
\end{tabular}
\end{center}
$^{a}$ Supplemented by $\delta_{\rm ov}$ = 1.7.
\end{table*}

The low-$l$ p-mode frequencies of each model are calculated by using the Guenther's
adiabatic oscillation code \citep{guen94}. For the modes with a given degree $l$, 
the frequencies $\nu_{\mathrm{corr}}(n)$ corrected from the near-surface effects 
of the model are computed by using equation \citep{kjel08}
\begin{equation}
\nu_{\mathrm{corr}}(n)=\nu_{\mathrm{mod}}(n)+a[\frac{\nu_{\mathrm{mod}}(n)}{\nu_{max}}]^{b},
\label{nucor}
\end{equation}
where $b$ is fixed to be 4.9, $a$ is a parameter, $\nu_{\mathrm{mod}}(n)$ is
the adiabatic oscillation frequencies of the model, $\nu_{\mathrm{max}}$ is 
the frequency of maximum power. The value of $\nu_{\mathrm{max}}$ is 
$1820$ $\mu$Hz. The value of $a$ is determined from the observed and
theoretical frequencies of the modes with the degree $l$ by using equations
(6) and (10) of \citet{kjel08}.

\subsection{Spectroscopic Constraints}

The effective temperature of KIC 11081729 may be 6440 K \citep{wrig03},
$6570^{+67}_{-75}$ K \citep{ammo06}, 6359 K \citep{pins12},
$6630\pm70$ K \citep{brun12}, or $6400\pm127$ K \citep{mol13}.
The value of [Fe/H] for KIC 11081729 is $0.16^{+0.14}_{-0.15}$ \citep{ammo06}.
Combining the value of $(Z/X)_{\odot}=0.023$ of the Sun given by \citet{grev98},
the ratio of surface metal abundance to hydrogen abundance,
$(Z/X)_{\mathrm{s}}$, is estimated to be approximately between 0.024
and 0.047. The value of [Fe/H] determined by \citet{brun12} is $-0.12\pm0.06$,
which corresponds to a $(Z/X)_{\mathrm{s}}$ approximately between 0.015
and 0.020; while that estimated by \citet{mol13} is $-0.19\pm0.22$,
i.e., the value of $(Z/X)_{\mathrm{s}}$ is approximately between 
0.009 and 0.025. \citet{mol13} have shown that the estimated atmospheric 
parameters of stars hotter than 6,000 K are affected by the method
of spectral analysis. Thus the atmospheric parameters determined by
\citet{ammo06}, \citet{brun12}, and \citet{mol13} were considered
in this work. 

The visual magnitude of KIC 11081729 is $9.048\pm0.054$ \citep{droe06, ammo06}.
The value of bolometric correction and extinction is estimated from the tables 
of \citet{flow96} and \citet{ammo06}, respectively. The distance of KIC 11081729
is $D=99^{+91}_{-36}$ pc \citep{ammo06, pick10}. Then the luminosity of KIC 11081729
is estimated to be $4.00\pm3.28$ $L_\odot$. The large uncertainty of the luminosity
derives from the uncertainty of the distance.

To find the models that can reproduce the observed characteristics of KIC 11081729, 
we calculated the value of $\chi_{\mathrm{c}}^{2}$ of models.
The function $\chi_{\mathrm{c}}^{2}$ is defined as
\begin{equation}
\chi^{2}_{\mathrm{c}} = \frac{1}{3}\sum_{i=1}^{3}
[\frac{C_{i}^{\mathrm{theo}}-C_{i}^{\mathrm{obs}}}{\sigma(C_{i}^{\mathrm{obs}})}]^{2},
\end{equation}
where the symbol $C = [L/L_{\odot}, T_{\rm eff}, (Z/X)_{s}]$, the $C_{i}^{\mathrm{obs}}$
presents the observed values of these parameters, while $C_{i}^{\mathrm{theo}}$ 
represents the theoretical values. The observational error is given by
$\sigma(C_{i}^{\mathrm{obs}})$. In the first step, the models with
$\chi_{\mathrm{c}}^{2}$ $\leq 1$ are chosen as candidate models.

\subsection{Asteroseismic Constraints}
In order to find the models that can reproduce the oscillation characteristics
of KIC 11081729, we computed the value of $\chi_{\nu}^{2}$ for each model.
The function $\chi_{\nu}^{2}$ is defined as
\begin{equation}
\chi_{\nu}^{2} = \frac{1}{N}\sum_{i=1}^{N}
[\frac{\nu_{i}^{\mathrm{theo}}-\nu_{i}^{\mathrm{obs}}}{\sigma(\nu_{i}^{\mathrm{obs}})}]^{2},
\end{equation}
where the quantity $\nu$ corresponds to the individual frequencies 
of modes from observation and modelling, $\sigma(\nu_{i}^{\mathrm{obs}})$ 
is the observational error of the $\nu_{i}^{\mathrm{obs}}$, $N$ presents the 
number of observed modes. The value of $\chi_{\nu_{\mathrm{corr}}}^{2}${} of 
corrected frequencies is also computed. The models with $\chi_{\mathrm{c}}^{2}$
$\leq1.0$ and $\chi_{\nu}^{2}$ or $\chi_{\nu_{\mathrm{corr}}}^{2}${} $<10.0$ 
are chosen as candidate models.

When a model evolves to the vicinity of the error-box
of $L/L_{\odot}$ and $T_{\rm eff}$ in the H-R diagram,
the evolutionary time step of the model is set as small as
possible to ensure that consecutive models have an approximately 
equal $\chi_{\nu_{\mathrm{corr}}}^{2}$. In some cases, there may be
a difference of a few Myr between the age of the model with 
the minimum $\chi_{\nu_{\mathrm{corr}}}^{2}$ and that of the model 
with the minimum $\chi_{\nu}^{2}$. \citet{yang15} showed that the difference 
between the model with the minimum $\chi_{\nu_{\mathrm{corr}}}^{2}${}
and the model with the minimum $\chi_{\nu}^{2}$ is insignificant.
For a given mass and $\delta_{\rm ov}$, the model with the minimum
$\chi_{\nu_{corr}}^{2}$ is chosen as the best model.

\subsubsection{The models with the effective temperature and [Fe/H] of Bruntt}

\begin{table*}
\begin{center}
\renewcommand\arraystretch{1.0}
\caption{Parameters of the models with the effective temperture and [Fe/H] determined
by \citet{brun12}. The symbol $X_{c}$ represents the central hydrogen abundance
of models, while $\nu^{m}_{0}$ indicates the adiabatic oscillation frequency
at which the $r_{01}$ of models reaches a minimum. 
\label{tab2}}
\begin{tabular}{p{0.70cm}cccccccccccccc}
  \hline\hline\noalign{\smallskip}
   Model & $M$ & $T_{\rm eff}$& $ L$ & $R$& age & $Z_{i}$ & $(Z/X)_{s}$ & $\alpha$ & $\delta_{\rm ov}$
   & $X_{c}$ & $\chi_{\nu}^{2}$ & $\chi_{\nu_{corr}}^{2}$ &  $\chi_{c}^{2}$ & $\nu^{m}_{0}$ \\

  & ($M_\odot$) & (K) & ($L_\odot$)  & ($R_\odot$) & (Gyr) & &  & & &  &  & & & ($\mu$Hz)\\
  \hline\hline\noalign{\smallskip}
 Mb1 & 1.21 & 6524 & 3.19 & 1.400 & 2.475 & 0.018 & 0.018 & 2.05 & 0.2 & 0.410 &  5.5 & 5.6 &  0.8 & 2255\\
 Mb2 & 1.22 & 6524 & 3.20 & 1.403 & 2.488 & 0.018 & 0.017 & 2.05 & 0.2 & 0.417 &  5.9 & 5.5 &  0.8 & 2255\\
 Mb3 & 1.23 & 6534 & 3.22 & 1.403 & 2.222 & 0.018 & 0.016 & 1.95 & 0.2 & 0.450 &  5.9 & 5.6 &  0.9 & 2346\\
 Mb4 & 1.24 & 6556 & 3.27 & 1.403 & 2.073 & 0.018 & 0.015 & 1.95 & 0.2 & 0.461 &  6.6 & 5.5 &  0.7 & 2347\\
 Mb5 & 1.25 & 6596 & 3.35 & 1.403 & 1.856 & 0.018 & 0.014 & 1.95 & 0.2 & 0.479 &  7.1 & 5.3 &  0.6 & 2347\\
 Mb6 & 1.26 & 6641 & 3.47 & 1.409 & 1.835 & 0.018 & 0.015 & 2.05 & 0.2 & 0.476 &  6.1 & 5.4 &  0.4 & 2347\\
 Mb7 & 1.27 & 6555 & 3.33 & 1.416 & 1.907 & 0.020 & 0.018 & 1.95 & 0.2 & 0.478 &  5.4 & 5.2 &  0.4 & 2347\\
 Mb8 & 1.28 & 6583 & 3.39 & 1.416 & 1.750 & 0.020 & 0.017 & 1.95 & 0.2 & 0.492 &  7.1 & 5.3 &  0.2 & 2347\\
 Mb9 & 1.29 & 6599 & 3.46 & 1.426 & 1.697 & 0.020 & 0.019 & 2.05 & 0.0 & 0.389 &  4.9 & 5.4 &  0.1 & 3562\\
 Mb10 &1.30 & 6553 & 3.40 & 1.432 & 2.452 & 0.016 & 0.022 & 1.95 & 0.0 & 0.327 &  5.5 & 5.5 &  0.9 & 4010\\
\noalign{\smallskip}\hline\hline
\end{tabular}
\end{center}
\end{table*}

With the constraints of the effective temperature
and the [Fe/H] of \citet{brun12}, we calculated the values of $\chi_{\mathrm{c}}^{2}$, $\chi_{\nu}^{2}$,
and $\chi_{\nu_{\mathrm{corr}}}^{2}${} of each model. Table \ref{tab2} lists the models that minimize
$\chi_{\mathrm{c}}^{2}$ + $\chi_{\nu_{\mathrm{corr}}}^{2}${} for a given mass. The value of $\delta_{\rm ov}$ of the models
is less than 0.4. The model Mb8 has the minimum $\chi_{\mathrm{c}}^{2}$ + $\chi_{\nu_{\mathrm{corr}}}^{2}${},
which seems to hint that Mb8 is the best model.

Using the posterior probability distribution function,
the mass estimated from the sample of the models with $\chi_{\mathrm{c}}^{2}$ $<1.0$
and $\chi_{\nu_{\mathrm{corr}}}^{2}${} $<10.0$ is $1.26\pm0.03$ $M_\odot$ for KIC 11081729,
where the error bar indicates the $68\%$ level confidence interval.
With the same constraints, \citet{metc14} also obtained
the same mass, i.e. $1.26\pm0.03$ $M_\odot$. The radius is 
estimated to be $1.41\pm0.01$ $R_\odot$ that is also close 
to $1.382\pm0.021$ $R_\odot$ of \citet{metc14}. However, the age 
of $1.9\pm0.4$ Gyr is larger than $0.86\pm0.21$ Gyr given by
\citet{metc14}. This is due to the fact that the effects of 
the settling of heavy elements and overshooting of convective 
core were considered that were not included in the models of
\citet{metc14}. This indicates that the estimated age of stars 
can be significantly affected by physical effects that are 
considered in models. 

\begin{figure}
\centering
\includegraphics[scale=0.34, angle=-90]{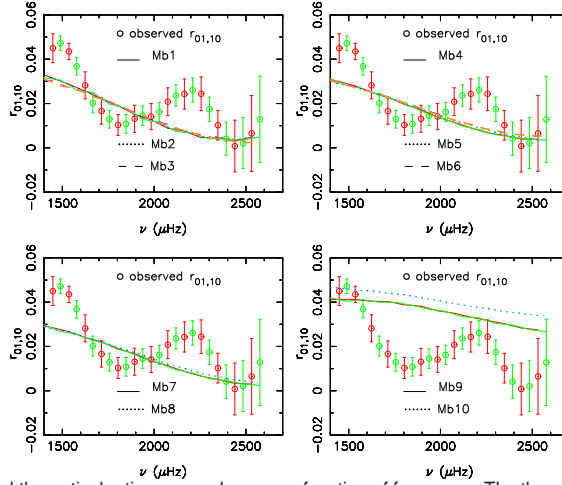}
\caption{The distributions of the observed and theoretical
ratios $r_{10}$ and $r_{01}$ as a function of frequency.
The theoretical ratios are computed from the corrected
oscillation frequencies $\nu_{\mathrm{corr}}$ of the models 
listed in Table \ref{tab2}.}
\label{pgb0}
\end{figure}

\begin{figure}
\centering
\includegraphics[scale=0.34, angle=-90]{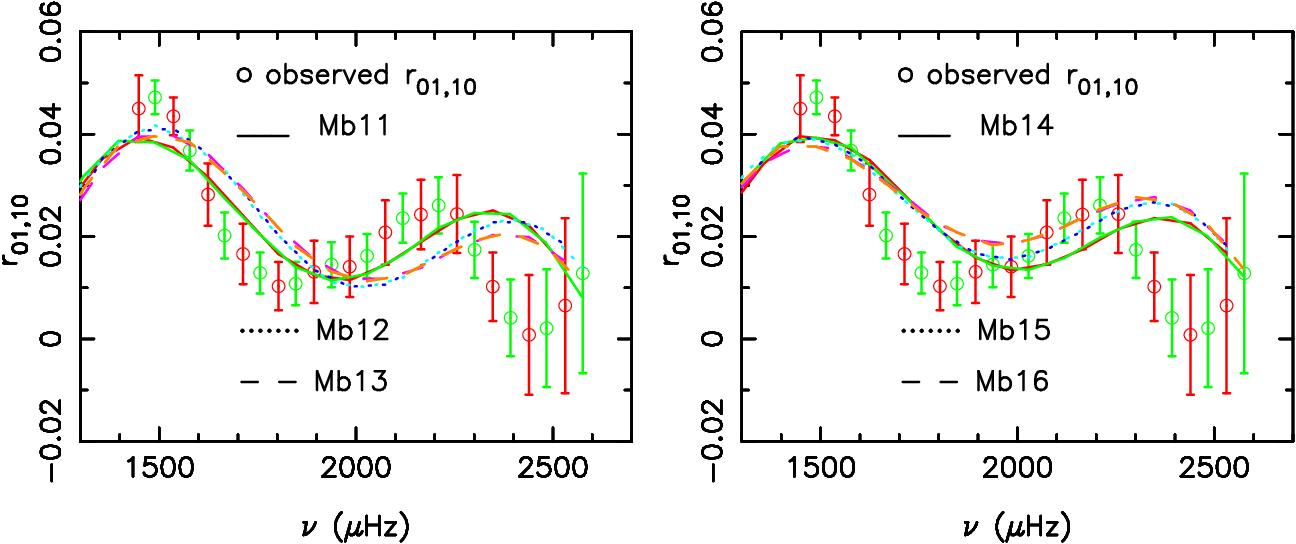}
\includegraphics[scale=0.34, angle=-90]{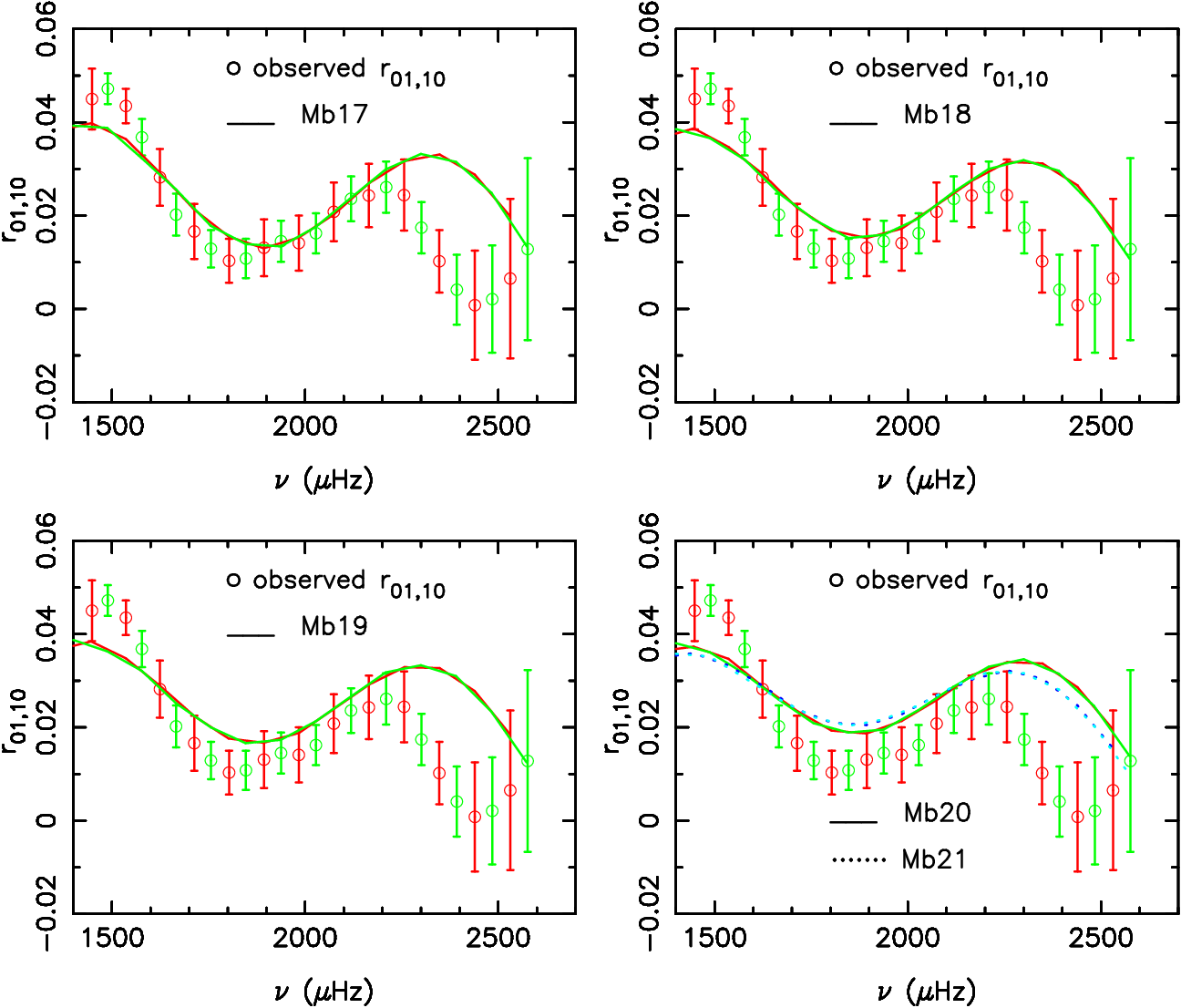}
\caption{The distributions of ratios $r_{10}$ and $r_{01}$
of the models listed in Table \ref{tab3}.}
\label{pgb8}
\end{figure}

However, the calculations show that the models with $\delta_{\rm ov}$ $\leq0.2$
can not reproduce the distributions of the observed $r_{10}$
and $r_{01}$ of KIC 11081729 (see Figure \ref{pgb0}).
This shows that the internal structures of these
models do not match that of KIC 11081729.
The observed ratios change dramatically at high frequencies,
which may derive from the fact that the observed frequencies
are affected by large-line widths at high frequencies
in the same way as that of KIC 6106415 and KIC 12009504 \citep{silv13}.
\citet{silv13} suggested that the ratios at high frequencies
should be excluded from the set of constraints.

\citet{liu14} argued that effects of overshooting of convective core
can lead to the fact that ratios $r_{10}$ and $r_{01}$ exhibit an
increase behavior. In order to test whether the characteristics
of $r_{10}$ and $r_{01}$ of KIC 11081729 can be reproduced by the 
effects of overshooting, we computed the evolutions of models
with $\delta_{\rm ov}$ as large as $1.8$. The calculation shows that 
models with $\delta_{\rm ov}$ $< 1.7$ can not reproduce the distributions 
of the observed $r_{10}$ and $r_{01}$. Table \ref{tab3} lists the
models that minimize $\chi_{\mathrm{c}}^{2}$ + $\chi_{\nu_{\mathrm{corr}}}^{2}${}
for a given mass and $\delta_{\rm ov}$. Model Mb19 has the minimum
$\chi_{\mathrm{c}}^{2}$ + $\chi_{\nu_{\mathrm{corr}}}^{2}${}.
Moreover, Figure \ref{pgb8} shows that the observed ratios
can be reproduced well by models with $\delta_{\rm ov}$ $=1.8$ and mass in
the range of $1.21-1.27$ $M_\odot$. 

\begin{table*}
\begin{center}
\renewcommand\arraystretch{1.0}
\caption{Parameters of the models with the effective temperture and [Fe/H] determined
by \citet{brun12}. The symbol $X_{c}$ represents the central hydrogen abundance
of models, while $\nu^{m}_{0}$ indicates the adiabatic oscillation frequency
at which the $r_{01}$ of models reaches a minimum. 
\label{tab3}}
\begin{tabular}{p{0.70cm}cccccccccccccc}
  \hline\hline\noalign{\smallskip}
   Model & $M$ & $T_{\rm eff}$& $ L$ & $R$& age & $Z_{i}$ & $(Z/X)_{s}$ & $\alpha$ & $\delta_{\rm ov}$
   & $X_{c}$ & $\chi_{\nu}^{2}$ & $\chi_{\nu_{corr}}^{2}$ &  $\chi_{c}^{2}$ & $\nu^{m}_{0}$ \\

  & ($M_\odot$) & (K) & ($L_\odot$)  & ($R_\odot$) & (Gyr) & &  & & &  &  & & & ($\mu$Hz)\\
  \hline\hline\noalign{\smallskip}
 Mb11 & 1.19 & 6560 & 3.19 & 1.384 & 4.415 & 0.020 & 0.015 & 1.95 & 1.7 & 0.566 & 20.8 & 8.9 &  0.8 &  818\\
 Mb12 & 1.21 & 6529 & 3.20 & 1.400 & 4.567 & 0.024 & 0.020 & 2.05 & 1.7 & 0.557 & 10.6 & 8.1 &  0.9 &  818\\
 Mb13 & 1.23 & 6605 & 3.36 & 1.400 & 3.793 & 0.022 & 0.017 & 2.05 & 1.7 & 0.574 & 13.7 & 8.0 &  0.1 &  817\\
 Mb14 & 1.25 & 6596 & 3.38 & 1.409 & 3.603 & 0.024 & 0.019 & 2.05 & 1.7 & 0.576 & 12.4 & 7.4 &  0.2 &  817\\
 Mb15 & 1.27 & 6585 & 3.40 & 1.419 & 3.426 & 0.026 & 0.022 & 2.05 & 1.7 & 0.579 &  9.8 & 7.0 &  0.7 &  817\\
 Mb16 & 1.29 & 6582 & 3.40 & 1.419 & 2.856 & 0.026 & 0.021 & 1.95 & 1.7 & 0.594 & 14.4 & 7.5 &  0.4 &  816\\
 \hline
 Mb17 & 1.21 & 6542 & 3.22 & 1.400 & 4.631 & 0.024 & 0.020 & 2.05 & 1.8 & 0.559 & 10.8 & 7.8 &  0.7 &  818\\
 Mb18 & 1.23 & 6536 & 3.22 & 1.403 & 4.000 & 0.024 & 0.019 & 1.95 & 1.8 & 0.574 & 14.9 & 7.7 &  0.6 &  817\\
 Mb19 & 1.25 & 6607 & 3.40 & 1.409 & 3.654 & 0.024 & 0.019 & 2.05 & 1.8 & 0.578 & 12.2 & 7.3 &  0.1 &  817\\
 Mb20 & 1.27 & 6640 & 3.49 & 1.413 & 3.210 & 0.024 & 0.018 & 2.05 & 1.8 & 0.588 & 14.0 & 7.5 &  0.1 &  816\\
 Mb21 & 1.29 & 6637 & 3.49 & 1.416 & 2.655 & 0.024 & 0.017 & 1.95 & 1.8 & 0.602 & 22.1 & 8.8 &  0.1 &  815\\
\noalign{\smallskip}\hline\hline
\end{tabular}
\end{center}
\end{table*}

Model Mb19 not only has the minimum $\chi_{\mathrm{c}}^{2}$ +
$\chi_{\nu_{\mathrm{corr}}}^{2}${} when $\delta_{\rm ov}$ is
in the range of $1.7-1.8$, it also reproduces the observed ratios.
This shows that the structure of model Mb19 is similar to 
that of KIC 11081729, and KIC 11081729 may have a large overshoot of
convective core. The mass, radius, and age of KIC 11081729
estimated from models with the large $\delta_{\rm ov}$ and 
with $\chi_{\mathrm{c}}^{2}$ $<1.0$ and $\chi_{\nu_{\mathrm{corr}}}^{2}${}
$<10.0$ are $1.24\pm0.03$ $M_\odot$,
$1.40\pm0.01$ $R_\odot$, and $3.6\pm0.7$ Gyr, respectively.
The mass and radius are close to those estimated by \citet{metc14}.
But the age is larger than that determined by \citet{metc14},
which is due to the fact that overshooting of convective core
brings more H-rich material into the region of hydrogen burning.

The value of $\chi_{\nu_{\mathrm{corr}}}^{2}${} of Mb19 is larger than that of Mb8.
Moreover, the effective temperature and [Fe/H] determined by \citet{mol13}
and \citet{ammo06} are different from those estimated by \citet{brun12}.
The determination of atmospheric parameters can be affected by spectroscopic
methods \citep{mol13}. Thus we calculated the models with the effective
temperature and [Fe/H] determined by \citet{mol13} and \citet{ammo06}.

\subsubsection{The models with the effective temperature and [Fe/H] of Molenda-Zakowicz}

\begin{table*}
\begin{center}
\renewcommand\arraystretch{1.0}
\caption{Parameters of the models with the effective temperture and [Fe/H] determined
by \citet{mol13}. The symbol $X_{c}$ represents the central hydrogen abundance
of models, while $\nu^{m}_{0}$ indicates the adiabatic oscillation frequency
at which the $r_{01}$ of models reaches a minimum.
\label{tab4} }
\begin{tabular}{p{0.70cm}cccccccccccccc}
  \hline\hline\noalign{\smallskip}
   Model & $M$ & $T_{\rm eff}$& $ L$ & $R$& age & $Z_{i}$ & $(Z/X)_{s}$ & $\alpha$ & $\delta_{\rm ov}$
   & $X_{c}$ & $\chi_{\nu}^{2}$ & $\chi_{\nu_{corr}}^{2}$ &  $\chi_{c}^{2}$ & $\nu^{m}_{0}$ \\

  & ($M_\odot$) & (K) & ($L_\odot$)  & ($R_\odot$) & (Gyr) & &  & & &  &  & & & ($\mu$Hz)\\
  \hline\hline\noalign{\smallskip}
 Mm1 & 1.17 & 6291 & 2.74 & 1.393 & 6.738 & 0.030 & 0.029 & 1.95 & 1.7 & 0.524 &  8.8 & 8.3 &  1.0 &  819\\
 Mm2 & 1.19 & 6326 & 2.82 & 1.400 & 6.070 & 0.030 & 0.028 & 1.95 & 1.7 & 0.533 &  8.5 & 8.0 &  0.8 &  818\\
 Mm3 & 1.21 & 6403 & 2.99 & 1.406 & 5.564 & 0.030 & 0.029 & 2.05 & 1.7 & 0.537 &  8.5 & 7.7 &  0.7 &  818\\
 Mm4 & 1.23 & 6476 & 3.15 & 1.413 & 4.707 & 0.028 & 0.026 & 2.05 & 1.7 & 0.553 &  8.3 & 7.6 &  0.5 &  818\\
 Mm5 & 1.25 & 6509 & 3.24 & 1.416 & 4.180 & 0.028 & 0.025 & 2.05 & 1.7 & 0.563 &  8.4 & 7.3 &  0.6 &  817\\
 Mm6 & 1.27 & 6543 & 3.33 & 1.421 & 3.706 & 0.028 & 0.025 & 2.05 & 1.7 & 0.572 &  8.4 & 6.9 &  0.8 &  817\\
 Mm7 & 1.29 & 6535 & 3.35 & 1.429 & 3.527 & 0.030 & 0.027 & 2.05 & 1.7 & 0.575 &  7.6 & 6.7 &  0.9 &  817\\
 \hline
  Mm8 & 1.17 & 6389 & 2.89 & 1.390 & 6.048 & 0.026 & 0.023 & 1.95 & 1.8 & 0.539 & 10.5 & 8.4 &  0.2 &  818\\
  Mm9 & 1.19 & 6423 & 2.98 & 1.396 & 5.421 & 0.026 & 0.023 & 1.95 & 1.8 & 0.549 & 10.6 & 8.2 &  0.2 &  818\\
 Mm10 & 1.21 & 6457 & 3.08 & 1.403 & 5.296 & 0.028 & 0.025 & 2.05 & 1.8 & 0.546 &  8.6 & 7.9 &  0.5 & 818\\
 Mm11 & 1.23 & 6489 & 3.17 & 1.409 & 4.773 & 0.028 & 0.025 & 2.05 & 1.8 & 0.556 &  8.5 & 7.8 &  0.5 & 818\\
 Mm12 & 1.25 & 6441 & 3.11 & 1.417 & 4.388 & 0.030 & 0.027 & 1.95 & 1.8 & 0.564 &  8.8 & 7.6 &  0.6 & 817\\
 Mm13 & 1.27 & 6515 & 3.26 & 1.419 & 3.607 & 0.028 & 0.024 & 1.95 & 1.8 & 0.579 & 11.0 & 7.1 &  0.5 &  816\\
 Mm14 & 1.29 & 6466 & 3.19 & 1.425 & 3.256 & 0.030 & 0.026 & 1.85 & 1.8 & 0.588 & 12.1 & 7.1 &  0.5 &  816\\
\noalign{\smallskip}\hline\hline
\end{tabular}
\end{center}
\end{table*}

\begin{figure}
\centering
\includegraphics[scale=0.34, angle=-90]{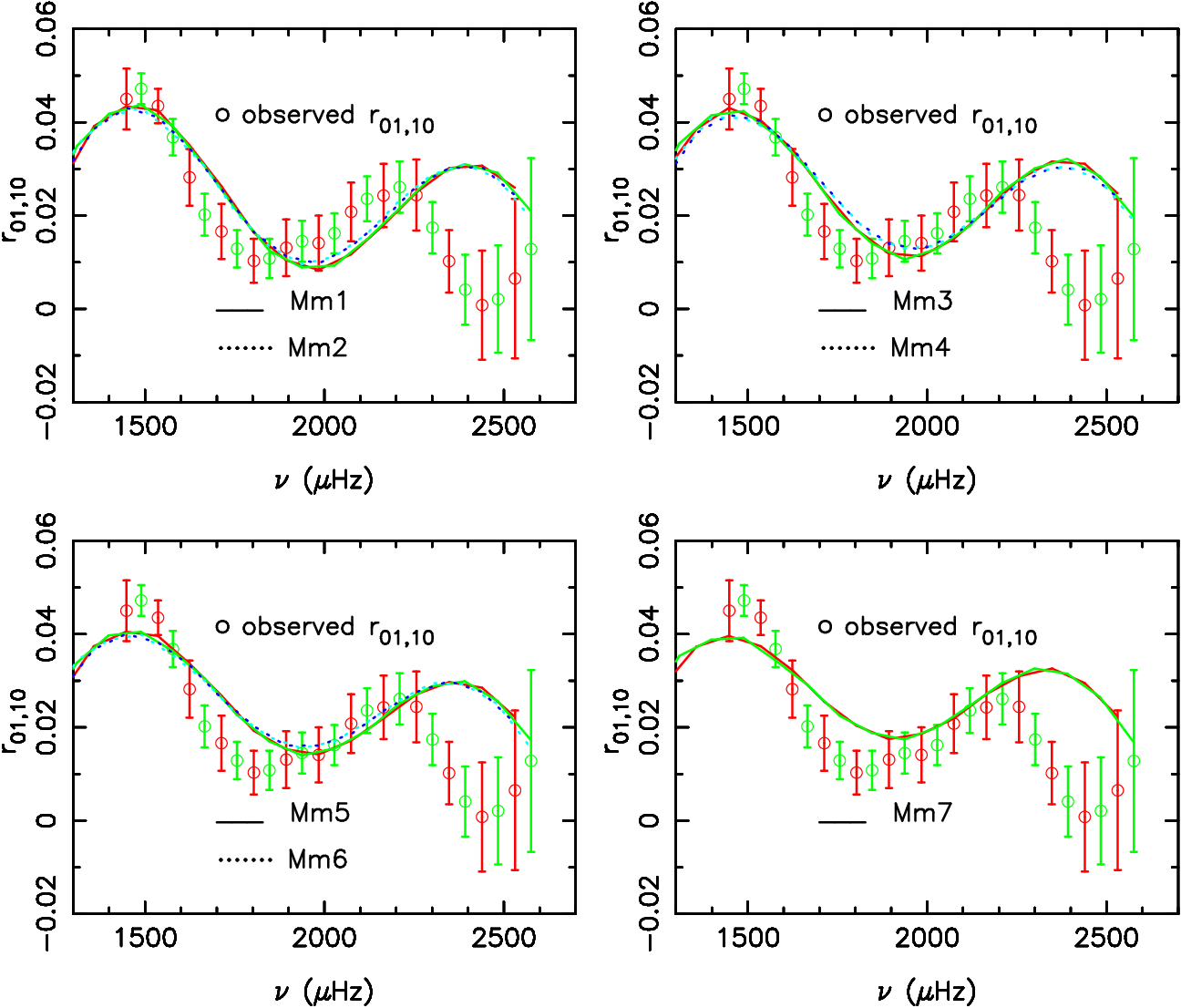}
\includegraphics[scale=0.34, angle=-90]{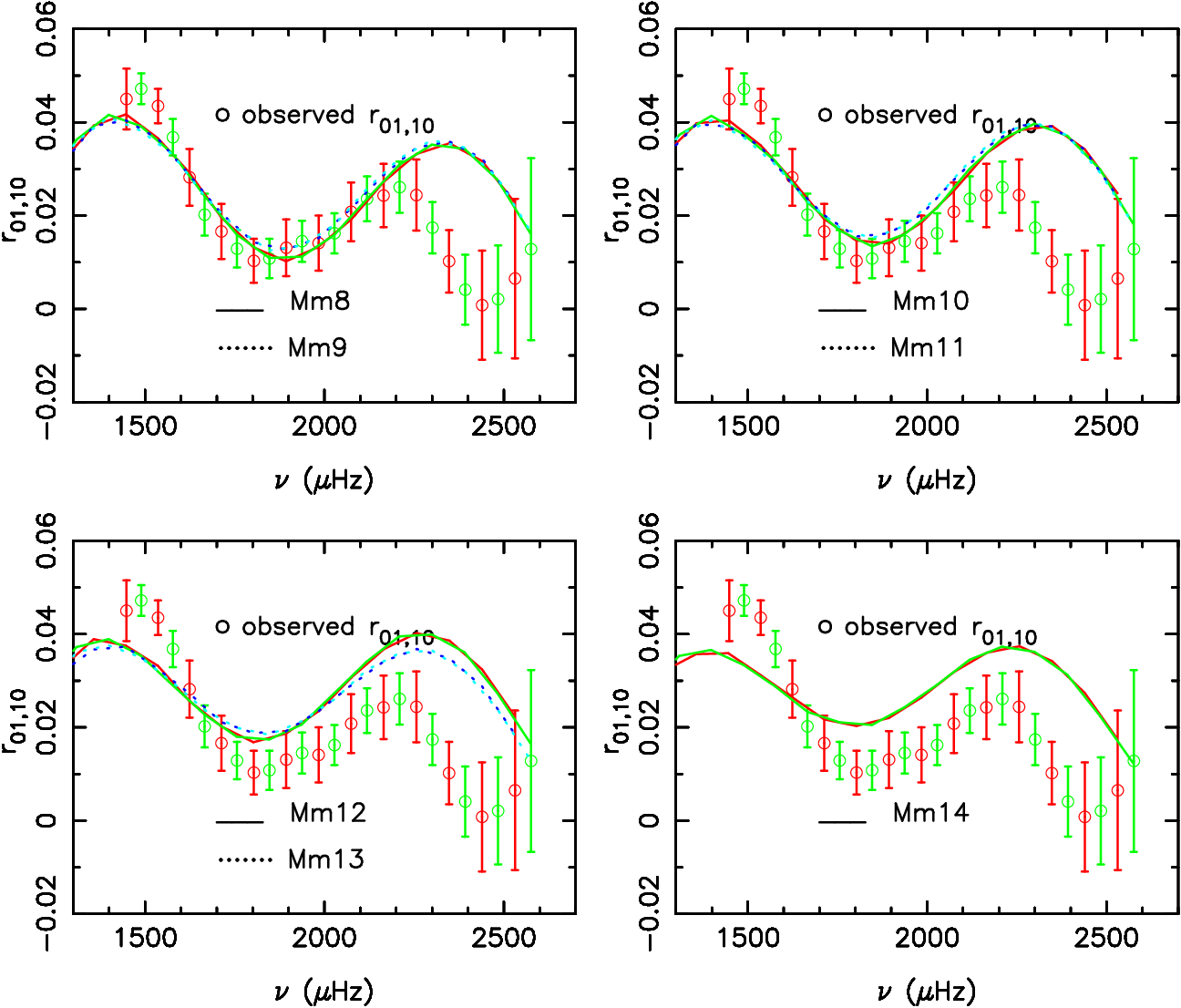}
\caption{The distributions of the observed and theoretical
ratios $r_{10}$ and $r_{01}$ as a function of frequency.
The theoretical ratios are computed from the corrected
frequencies $\nu_{\mathrm{corr}}$ of the models listed in Table \ref{tab4}.}
\label{pgm8}
\end{figure}

Table \ref{tab4} lists the models that minimize $\chi_{\mathrm{c}}^{2}$ 
+ $\chi_{\nu_{\mathrm{corr}}}^{2}${} for a given mass and $\delta_{\rm ov}$. 
The calculations show that the models
with $\delta_{\rm ov}$ $< 1.7$ can not reproduce the observed ratios.
Thus we do not considere these models when we estimate the
mass and age of KIC 11081729. When the value of $\delta_{\rm ov}$
increases to 1.8, the observed ratios can be reproduced
by models (see Figure \ref{pgm8}). This indicates that
the distributions of ratios $r_{10}$ and $r_{01}$ are mainly
dependent on the effects of overshooting of convective core
rather than on other effects.

The mass, radius, and age estimated from the models with $\delta_{\rm ov}$ in the
range of $1.7-1.8$ and with $\chi_{\mathrm{c}}^{2}$ $<1.0$ and $\chi_{\nu_{\mathrm{corr}}}^{2}${}
$<10.0$ are $1.230\pm0.035$ $M_\odot$, $1.40\pm0.01$ $R_\odot$,
and $4.6\pm0.9$ Gyr, respectively. The age of these models are
larger than that of models with the effective temperature
and [Fe/H] of \citet{brun12}. Thus spectroscopic results can
affect the estimation of age of stars. 
The values of $\chi_{\nu_{\mathrm{corr}}}^{2}${} of these models are as large as those 
of the models with the effective temperature and [Fe/H] 
of \citet{brun12}. Thus these models are not better
than the models with the effective temperature
and [Fe/H] of \citet{brun12}.

\subsubsection{The models with the effective temperature and [Fe/H] of Ammons }
\begin{figure}
\centering
\includegraphics[scale=0.34, angle=-90]{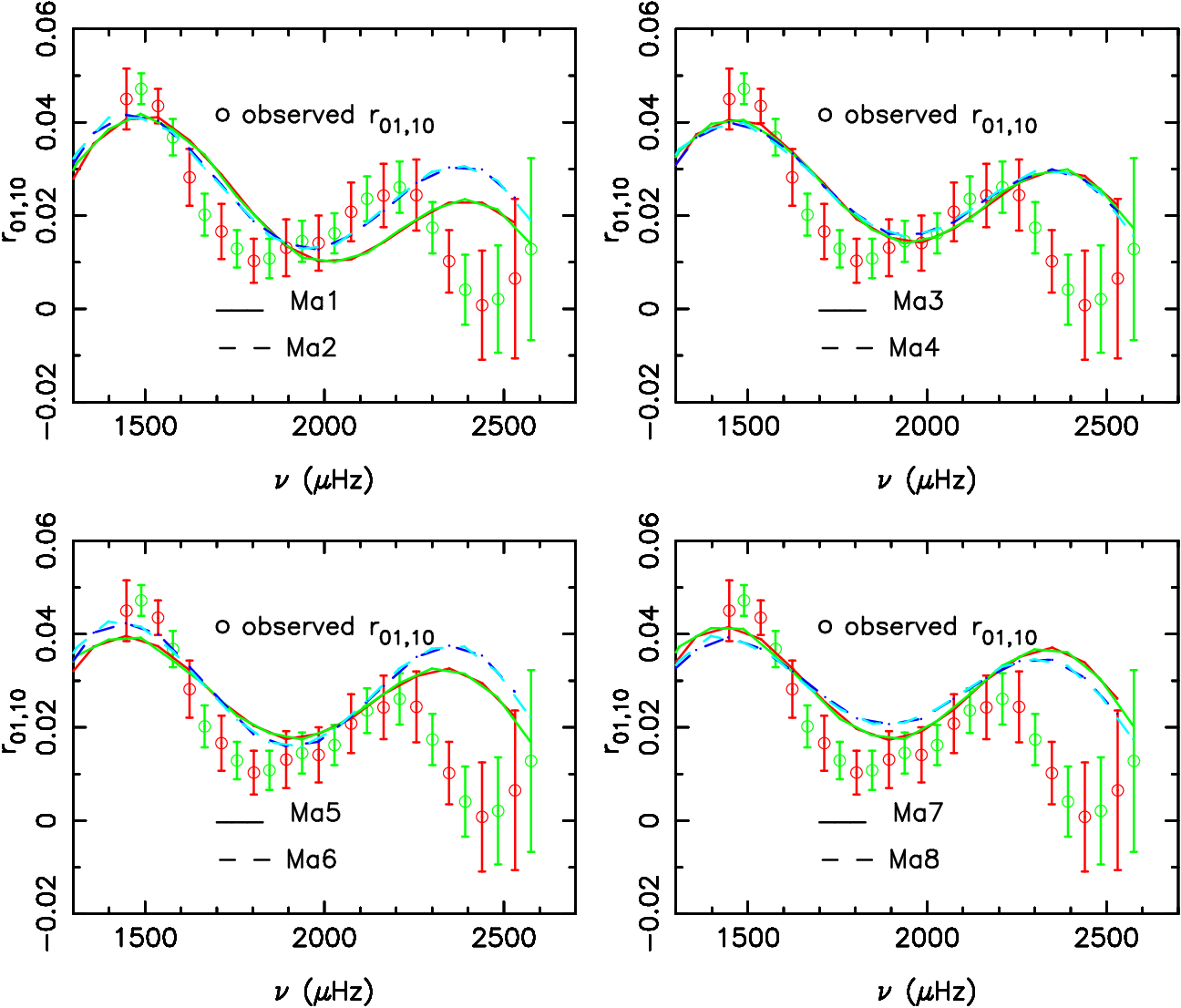}
\includegraphics[scale=0.34, angle=-90]{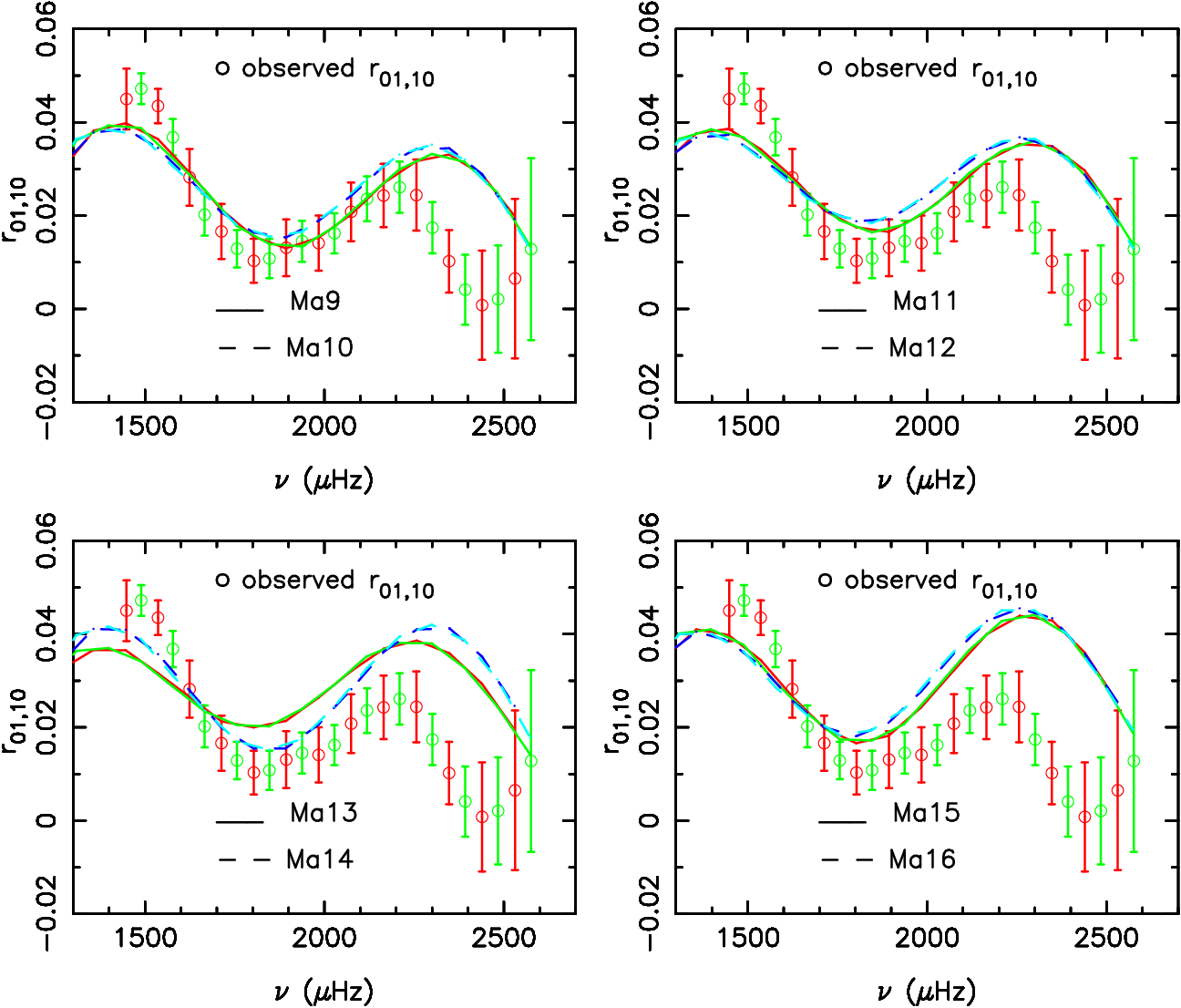}
\caption{The distributions of the observed and theoretical
ratios $r_{10}$ and $r_{01}$ as a function of frequency.
The theoretical ratios are computed from the corrected
frequencies $\nu_{\mathrm{corr}}$ of the models listed in Table \ref{tab5}.}
\label{pga8}
\end{figure}

The effective temperature determined by \citet{ammo06} is between that estimated 
by \citet{mol13} and that determined by \citet{brun12}. Table \ref{tab5} lists
the models that minimize $\chi_{\mathrm{c}}^{2}$ + $\chi_{\nu_{\mathrm{corr}}}^{2}${}
for a given mass and $\delta_{\rm ov}$. The distributions of $r_{10}$ and $r_{01}$ of
models with $\delta_{\rm ov}$ less than 1.7 are also not consistent with those of the
observed ratios. When the value of $\delta_{\rm ov}$ is in the range of $1.7-1.8$,
the observed ratios are reproduced well by the models with the
large $\delta_{\rm ov}$ (see Figure \ref{pga8}).

\begin{table*}
\begin{center}
\renewcommand\arraystretch{1.0}
\caption{Parameters of the models with the effective temperture and [Fe/H] determined
by \citet{ammo06}. The symbol $X_{c}$ represents the central hydrogen abundance
of models, while $\nu^{m}_{0}$ indicates the adiabatic oscillation frequency
at which the $r_{01}$ of models reaches a minimum. 
\label{tab5}}
\begin{tabular}{p{0.70cm}cccccccccccccc}
  \hline\hline\noalign{\smallskip}
   Model & $M$ & $T_{\rm eff}$& $ L$ & $R$& age & $Z_{i}$ & $(Z/X)_{s}$ & $\alpha$ & $\delta_{\rm ov}$
   & $X_{c}$ & $\chi_{\nu}^{2}$ & $\chi_{\nu_{corr}}^{2}$ &  $\chi_{c}^{2}$ & $\nu^{m}_{0}$ \\

  & ($M_\odot$) & (K) & ($L_\odot$)  & ($R_\odot$) & (Gyr) & &  & & &  &  & & & ($\mu$Hz)\\
  \hline\hline\noalign{\smallskip}
 Ma1 & 1.21 & 6529 & 3.20 & 1.400 & 4.567 & 0.024 & 0.020 & 2.05 & 1.7 & 0.557 & 10.6 & 8.1 &  0.7 &  818\\
 Ma2 & 1.23 & 6476 & 3.15 & 1.413 & 4.707 & 0.028 & 0.026 & 2.05 & 1.7 & 0.553 &  8.3 & 7.6 &  0.9 &  818\\
 Ma3 & 1.25 & 6509 & 3.24 & 1.416 & 4.180 & 0.028 & 0.025 & 2.05 & 1.7 & 0.563 &  8.4 & 7.3 &  0.5 &  817\\
 Ma4 & 1.27 & 6543 & 3.33 & 1.421 & 3.706 & 0.028 & 0.025 & 2.05 & 1.7 & 0.572 &  8.4 & 6.9 &  0.4 &  817\\
 Ma5 & 1.29 & 6536 & 3.35 & 1.429 & 3.527 & 0.030 & 0.027 & 2.05 & 1.7 & 0.575 &  7.6 & 6.7 &  0.3 &  817\\
 Ma6 & 1.30 & 6597 & 3.52 & 1.439 & 4.209 & 0.034 & 0.051 & 2.15 & 1.7 & 0.554 &  5.5 & 4.8 &  0.7 &  814\\
 Ma7 & 1.32 & 6630 & 3.61 & 1.442 & 3.735 & 0.034 & 0.051 & 2.15 & 1.7 & 0.563 &  5.1 & 4.6 &  0.9 &  814\\
 Ma8 & 1.34 & 6592 & 3.54 & 1.444 & 2.939 & 0.034 & 0.051 & 1.95 & 1.7 & 0.583 & 10.9 & 6.4 &  0.7 &  814\\
 \hline
  Ma9 & 1.21 & 6542 & 3.22 & 1.398 & 4.631 & 0.024 & 0.020 & 2.05 & 1.8 & 0.559 & 10.8 & 7.8 &  0.7 &  818\\
 Ma10 & 1.23 & 6490 & 3.15 & 1.406 & 4.296 & 0.026 & 0.022 & 1.95 & 1.8 & 0.568 & 11.6 & 7.7 &  0.9 &  817\\
 Ma11 & 1.25 & 6562 & 3.32 & 1.413 & 3.934 & 0.026 & 0.022 & 2.05 & 1.8 & 0.572 & 10.0 & 7.2 &  0.5 &  817\\
 Ma12 & 1.27 & 6515 & 3.26 & 1.419 & 3.607 & 0.028 & 0.024 & 1.95 & 1.8 & 0.579 & 11.0 & 7.1 &  0.6 &  816\\
 Ma13 & 1.29 & 6507 & 3.28 & 1.426 & 3.424 & 0.030 & 0.026 & 1.95 & 1.8 & 0.582 &  9.0 & 7.0 &  0.5 &  816\\
 Ma14 & 1.30 & 6504 & 3.31 & 1.435 & 5.298 & 0.026 & 0.036 & 1.95 & 1.8 & 0.599 &  7.4 & 5.7 &  0.3 &  815\\
 Ma15 & 1.32 & 6487 & 3.32 & 1.443 & 5.054 & 0.028 & 0.039 & 1.95 & 1.8 & 0.602 &  6.5 & 5.8 &  0.5 &  815\\
 Ma16 & 1.34 & 6476 & 3.33 & 1.451 & 4.724 & 0.030 & 0.042 & 1.95 & 1.8 & 0.604 &  6.0 & 5.8 &  0.7 &  814\\
 \hline
 Ma11b &1.25 & 6522 & 3.26 & 1.415 & 4.240 & 0.028 & 0.025 & 2.05 & 1.8 & 0.565 &  8.5 & 7.4 &  0.5 &  817\\
\noalign{\smallskip}\hline\hline
\end{tabular}
\end{center}
\end{table*}

Tables \ref{tab1} $-$ \ref{tab5} show that model Ma6 has the minimum
$\chi_{\mathrm{c}}^{2}$ + $\chi_{\nu_{\mathrm{corr}}}^{2}${} in the calculations. 
Moreover, the observed ratios are reproduced
by Ma6. Thus Ma6 is the best-fit model in the calculations.
When the value of $\delta_{\rm ov}$ is equal to 1.8, models Ma14
has the minimum $\chi_{\mathrm{c}}^{2}$ + $\chi_{\nu_{\mathrm{corr}}}^{2}${}. Figure \ref{pga8}
shows that the ratios of model Ma14 is almost consistent with
the observed ones. Therefore, Ma14 is chosen as the best-fit model
for $\delta_{\rm ov}$ = 1.8.

\begin{figure}
\centering
\includegraphics[scale=0.5, angle=-90]{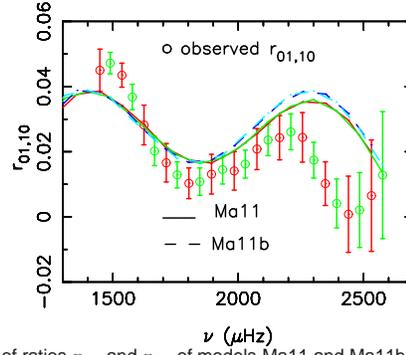}
\caption{The distributions of ratios $r_{10}$ and $r_{01}$ of models Ma11 and Ma11b
as a function of frequency.}
\label{pgcom}
\end{figure}

The mass, radius, and age estimated from the models with
$\delta_{\rm ov}$ in the range of $1.7-1.8$ and with $\chi_{\mathrm{c}}^{2}$ $<1.0$
and $\chi_{\nu_{\mathrm{corr}}}^{2}${} $<10.0$ are $1.270\pm0.035$ $M_\odot$,
$1.420\pm0.015$ $R_\odot$, and $3.8\pm0.7$ Gyr, respectively.
The median age of $3.8$ Gyr is larger than the ages of models with
$M=1.27$ $M_\odot$ listed in Table \ref{tab5}. This is related to
the fact that the table only lists part of the sample. 
For example, models Ma11 and Ma11b have the same mass and $\chi_{\mathrm{c}}^{2}$, 
and almost the same $\chi_{\nu_{\mathrm{corr}}}^{2}${}, $r_{10}$ and $r_{01}$ (see Figure \ref{pgcom}),
however, they have different age. 

The calculations show that the estimated mass and age rely
on spectroscopic results but the distributions of $r_{10}$ and $r_{01}$
are mainly dependent on the effects of overshooting of convective core.
The estimation of $\delta_{\rm ov}$ is not affected by the difference in the 
spectroscopic results. KIC 11081729 may have a large overshooting
of the convective core.

\section{Fitting equation for the observed and theoretical ratios}
The distributions of $r_{10}$ and $r_{01}$ are dependent on $\delta_{\rm ov}$,
which may provide an opportunity to diagnose the size of convective
core from observed frequencies.

\begin{figure}
\centering
\includegraphics[scale=0.5, angle=-90]{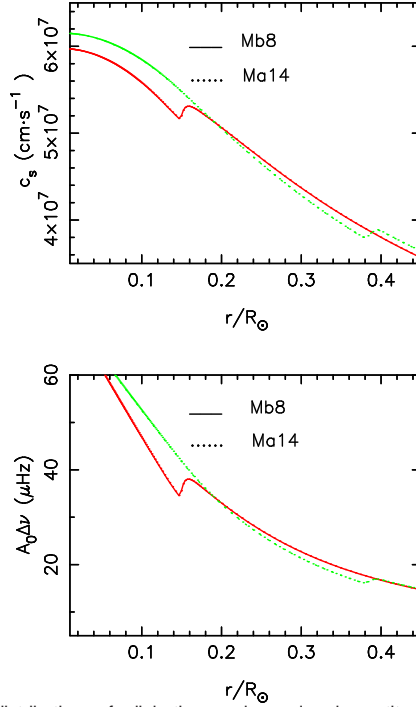}
\caption{ Radial distributions of adiabatic sound speed and quantity
$A_{0}\Delta\nu$ of models. }
\label{pgc0}
\end{figure}

By making use of the asymptotic formula of frequencies, one can get \citep{liu14}
\begin{equation}
r_{10}(\nu_{n,1}) \simeq 2\nu^{-1}_{n,1}A_{0}\Delta\nu,
\label{r1}
\end{equation}
where
\begin{equation}
A_{0}\Delta\nu=\frac{1}{4\mathrm{\pi}^2}[\frac{c(R)}{R}+\int_{r_{t}}^{R}(-\frac{1}{r}\frac{dc}{dr})dr].
\label{av0}
\end{equation}
In this equation, $c$ is the adiabatic sound speed at radius $r$ and $R$
is the fiducial radius of the star; $r_{t}$ is the inner turning point
of the mode with the frequency $\nu_{n,1}$. Equation (\ref{r1}) indicates
that ratio $r_{10}$ is dependent on the quantity $A_{0}\Delta\nu$ that is
sensitive to the changes in the adiabatic sound speed $c$. Figure \ref{pgc0}
shows that the changes can directly affect the quantity $A_{0}\Delta\nu$. 
These hint to us that ratios $r_{10}$ and $r_{01}$ of stars with a convective
core could be affected by the changes in the sound speed.

The p-mode oscillations of stars are considered to be acoustic 
standing waves. The variation with time $t$ of a standing wave is
proportional to $\cos(\omega t)$, where the $\omega$ is angular frequency. 
The variation of the modes that penetrate into overshooting region can
be affected by convection. For $l=1$ modes, there should be a critical 
angular frequency $\omega_{0}$. When the angular frequency of modes 
is larger than $\omega_{0}$, the modes penetrate into overshooting region.
We assume that the variation with time of the acoustic wave that is 
affected by convection is related to $-A\cos(\omega_{0}t)$,
where $A$ is a free parameter. Due to the fact that oscillation 
frequencies of stars can be observed, thus we want to know
that the effects of convection core on the characteristics of 
oscillations vary with frequency rather than with time.  
This can be achieved by making use of Fourier transform.
Thus one can obtain 
\begin{equation}
  \begin{array}{ll}
B(\omega)&=\frac{2}{\mathrm{\pi}}\int^{2N\mathrm{\pi}/\omega_{0}}_{0}[-A\cos(\omega_{0}t)\cos(\omega t)]dt \\
         &=-\frac{A}{\mathrm{\pi}}[\frac{\sin(\frac{2N\mathrm{\pi}\omega}{\omega_{0}})}{\omega+\omega_{0}}
            +\frac{\sin(\frac{2N\mathrm{\pi}\omega}{\omega_{0}})}{\omega-\omega_{0}}]  \\
         &=\frac{2A\omega}{\mathrm{\pi}(\omega^{2}_{0}-\omega^{2})}\sin(\frac{2N\mathrm{\pi}\omega}{\omega_{0}}),\\
\end{array}
\label{o1}
\end{equation}
where $N$ is an integer. Taken $N$ as $1$ and $\omega_{n,1}=2\pi\nu_{n,1}$, 
equation (\ref{o1}) can be rewritten as 
\begin{equation}
B(\nu_{n,1})=\frac{2A\nu_{n,1}}{2\mathrm{\pi^{2}}(\nu^{2}_{0}-\nu_{n,1}^{2})}
          \sin(\frac{2\mathrm{\pi}\nu_{n,1}}{\nu_{0}})+B_{0},
\label{o2}
\end{equation}
where $B_{0}$ is a constant. The effects of convection core on the 
characteristics of oscillations should be shown by function $B(\nu_{n,1})$.

The value of $B(\nu_{n,1})$ reaches maxima at around $3\nu_{0}/8$ 
and $7\nu_{0}/4$ , but reaches minima at around $\nu_{0}$ and $9\nu_{0}/4$.
Thus the value of $B(\nu_{n,1})$ decreases with frequency between
about $3\nu_{0}/8$ and $\nu_{0}$, but increases with frequency
between about $\nu_{0}$ and $7\nu_{0}/4$. The value of $B(\nu_{n,1})$
decreases firstly and then increases with frequency in the range of
about $7\nu_{0}/4 - 11\nu_{0}/4$. These characteristics are
consistent with those of $r_{10}$ and $r_{01}$ calculated
from models with a convective core. Thus the ratios $r_{10}$ 
and $r_{01}$ of stars with a convective core could be described
by equation (\ref{o2}). 

\begin{figure}
\centering
\includegraphics[scale=0.6, angle=-90]{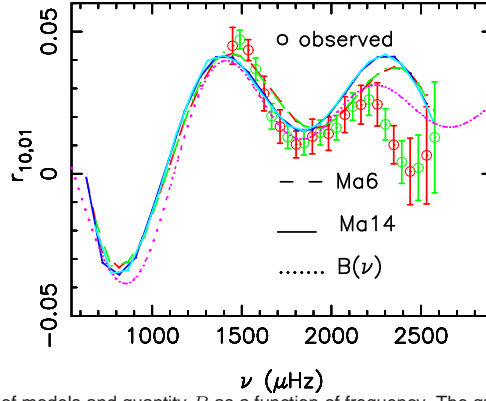}
\caption{The distributions of ratios $r_{10}$ and $r_{01}$ of models
and quantity $B$ as a function of frequency. The quantity $B$ is computed
by using equation (\ref{o2}) with $A=50\pi$, $B_{0}=0.023$, 
and $\nu_{0}=820$ $\mu$Hz}.
\label{pgsin}
\end{figure}

The observed ratios of KIC 11081729 decrease firstly and then increase
with frequency and have a minimum between 1804 and 1846 $\mu$Hz. Thus 
one can estimate the value of $\nu_{0}$ is between about 800 and 820 $\mu$Hz.
Due to the fact that the value of $\sin(2\pi \nu_{n,1}/\nu_{0})$ is $0$
at $\nu_{0}/2$ and $3\nu_{0}/2$, hence $B_{0}$ is not a free parameter.
For a given $\nu_{0}$, the value of $B_{0}$ takes the value of ratio $r_{10}$ 
or $r_{01}$ at $\nu_{0}/2$ or $3\nu_{0}/2$. Thus we obtained the value of
$B_{0}$ is between 0.019 and 0.026 from the ratios of Ma6. In addition, using 
equation (\ref{o2}) and the observed $\nu_{n,1}$ and $r_{10}$,
the values of parameters $A$, $B_{0}$, and $\nu_{0}$ are estimated
to be $53\pm21$ $\pi$, $0.018\pm0.003$, and $795\pm21$ $\mu$Hz,respectively.
Figure \ref{pgsin} shows that the ratios of KIC 11081729 are 
reproduced well by equation (\ref{o2}) with $A=50 \pi$, $B_{0}=0.023$, 
and $\nu_{0}=820$ $\mu$Hz. The distributions of $r_{10}$ and $r_{01}$ of Ma6 
and Ma14 are reproduced well by equation (\ref{o2}). The ratios $r_{10}$ 
and $r_{01}$ of models Ma6 and Ma14 reach the minimum at about $810$ $\mu$Hz 
that are consistent with $795\pm21$ $\mu$Hz estimated by using 
equation (\ref{o2}).

\subsection{Test the equation by other stars}

\begin{figure}
\centering
\includegraphics[scale=0.4, angle=-90]{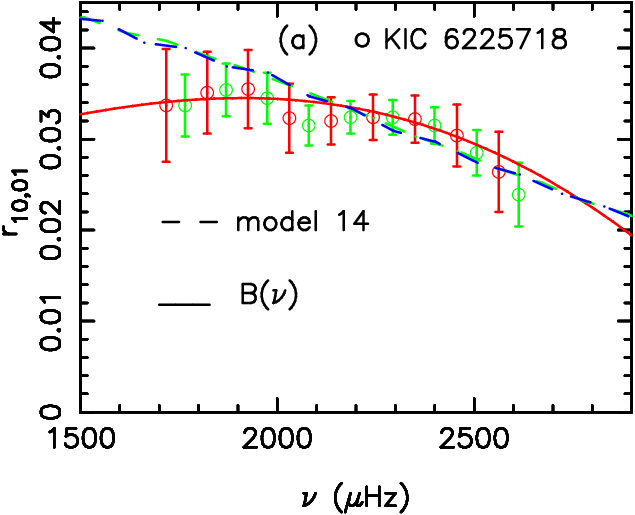}
\includegraphics[scale=0.4, angle=-90]{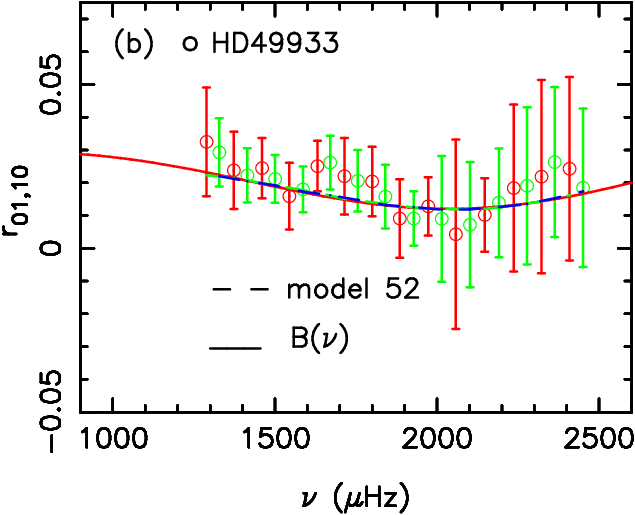}
\includegraphics[scale=0.4, angle=-90]{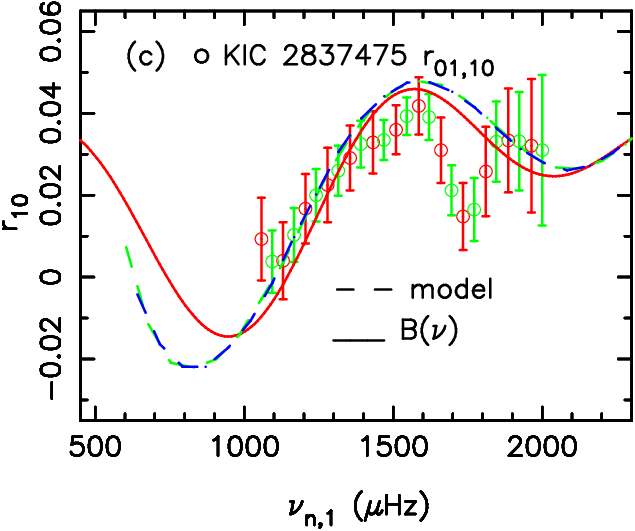}
\caption{The distributions of $r_{10}$ and $r_{01}$ of different
stars as a function of frequency.  }
\label{pg499}
\end{figure}

\citet{tian14} studied the oscillations of KIC 6225718, and determined 
that the mass of KIC 6225718 is $1.10^{+0.04}_{-0.03}$ $M_\odot$,
but could not determine whether a convective core exists in KIC 6225718.
From observed frequencies, one can obtain $A=745\pm527$ $\pi$, 
$B_{0}=0.021\pm0.008$, and $\nu_{0}=5700\pm423$ $\mu$Hz for KIC 6225718.
The panel (a) of Figure \ref{pg499} shows that the observed ratios of
KIC 6225718 are reproduced well by equation (\ref{o2}) with $A=745$ $\pi$, 
$B_{0}=0.021$, and $\nu_{0}=5700$ $\mu$Hz. Using equation (16) of
\citet{liu14} with $f_{0}=2$, the value of $\nu_{0}$ for the 
best model 14 of KIC 6225718 \citep{tian14} is estimated to 
be $5890$ $\mu$Hz. According to equation (\ref{o2}), ratios 
$r_{10}$ and $r_{01}$ reach the minimum at around $\nu_{0}$. 
The ratios $r_{10}$ and $r_{01}$ calculated from the model 14  
reach the minimum at $5642$ $\mu$Hz. The two values 
are consistent with $5700\pm423$ $\mu$Hz. The ratios 
of KIC 6225718 mainly exhibit a decreasing behavior. 
But the observed ratios have a maximum at about $1926$ $\mu$Hz, 
which is consistent with that $B(\nu)$ has a maximum at around
$3\nu_{0}/8$. The model 14 has a small convective core 
and $\delta_{\rm ov}$ $=0.2$ \citep{tian14}. These indicate
that a small convective core may exist in KIC 6225718.
The value of $\nu_{\mathrm{max}}$ of KIC 6225718 is $2031$ $\mu$Hz
that is much less than the value of $\nu_{0}$ ($5700\pm423$ $\mu$Hz).

The values of $A$, $B_{0}$, and $\nu_{0}$ computed from the observed 
frequencies of HD 49933 \citep{beno09} are $54\pm14$ $\pi$, $0.040\pm0.006$, 
and $1920\pm46$ $\mu$Hz. Using equation (16) of \citet{liu14} with $f_{0}=2$, 
the value of $\nu_{0}$ for model M52 of HD 49933 \citep{liu14} 
is estimated to be about $1900$ $\mu$Hz. The ratios of model M52 reach 
the minimum at about $2057$ $\mu$Hz. The panel (b) of Figure \ref{pg499} 
shows that the ratios of model M52 are consistent with those 
computed by equation (\ref{o2}) with $A=31$ $\pi$, $B_{0}=0.028$, 
and $\nu_{0}=1966$ $\mu$Hz. The model M52 has a large convective
core and $\delta_{\rm ov}$ $=0.7$ \citep{liu14}. The value of 
$\nu_{\mathrm{max}}$ of HD 49933 is $1760$ $\mu$Hz \citep{appo08}
that is close to $1920\pm46$ $\mu$Hz of $\nu_{0}$. 
The values of extracted frequencies of HD 49933 
are in the vicinity of $\nu_{0}$. The ratios of HD 49933 
decrease and then increase with frequency, which is consistent with 
that calculated from equation (\ref{o2}).

Moreover, the panel (c) of Figure \ref{pg499} shows that 
the observed and theoretical ratios of KIC 2837475 are reproduced
by equation  (\ref{o2}) with $A=43$ $\pi$, $B_{0}=0.033$, 
and $\nu_{0}=913$ $\mu$Hz in the range of about $1000-2000$ $\mu$Hz.
The values of $\nu_{0}$ computed from the observed frequencies 
of KIC 2837475 is $913\pm36$ $\mu$Hz. The ratios $r_{10}$ and $r_{01}$ 
calculated from model Ma14 of KIC 2837475 \citep{yang15} reach 
the minimum at about $850$ $\mu$Hz. The model has a large convective
core and $\delta_{\rm ov}$ $=1.4$ \citep{yang15}. 
The value of $\nu_{\mathrm{max}}$ of KIC 2837475 is $1522$ $\mu$Hz
that is much larger than $913\pm36$ $\mu$Hz
of $\nu_{0}$. The values of extracted frequencies of KIC 2837475
are in the range between about $\nu_{0}$ and $9\nu_{0}/4$. 
The ratios of KIC 2837475 increase and then decrease with frequency, 
which is also consistent with that calculated from equation (\ref{o2}).

The distributions of $r_{10}$ and $r_{01}$ of different stars with 
a convective core can be reproduced well by equation (\ref{o2}).
The value of $\nu_{\mathrm{max}}$ of KIC 6225718 is less than the value
of $\nu_{0}$. KIC 6225718 may have a small convective core and 
a small $\delta_{\rm ov}$. The value of $\nu_{\mathrm{max}}$ of HD 49933 
is close to the value of $\nu_{0}$. HD 49933 may have a medium $\delta_{\rm ov}$.
The value of $\nu_{\mathrm{max}}$ of KIC 2837475 and KIC 11081729 is larger
than that of $\nu_{0}$, respectively. KIC 2837475 and KIC 11081729 
may have a large $\delta_{\rm ov}$. The equation 
(\ref{o2}) could be used to determine the presence of
convective core and estimate the size of overshoot
from observed frequencies.

\section{DISCUSSION AND SUMMARY}

\subsection{Discussion}

\citet{mazu14} showed that ratios $r_{10}$ and $r_{01}$ can be affected
by the glitches at the BCZ and the layers of the HeIIZ. 
The changes in frequencies caused by the glitch at the BCZ
or the layers of the HeIIZ have a periodicity of twice the acoustic
depth of the corresponding glitch \citep{mazu14}. The value of
the acoustic depth $\tau_{HeIIZ}$ and $\tau_{BCZ}$ is about
720 and 2060 s for model Mb8, respectively. The value of $\tau_{HeIIZ}$
and $\tau_{BCZ}$ of model Mb19 is 728 and 2100 s, respectively.
The values of $\tau_{HeIIZ}$ and $\tau_{BCZ}$ of model Mb8 are
approximately equal to those of model Mb19. Thus the glitches
should result in a similar effect on the ratios of models Mb8 
and Mb19. However, the distributions of ratios $r_{10}$ and $r_{01}$ 
of model Mb8 are obviously different from those of model Mb19. 
Thus, the difference between the ratios of Mb8 and those of Mb19
(see Figures \ref{pgb0} and \ref{pgb8}) does not derive from 
the effects of the glitch at the BCZ or the HeIIZ.

For different spectroscopic results, only the models with $\delta_{\rm ov}$ in the
range of about $1.7-1.8$ can reproduce the observed ratios.
This indicates that the uncertainties of the effective temperature
and [Fe/H] can not significantly affect the estimation of $\delta_{\rm ov}$.
This is due to the fact that the estimation of $\delta_{\rm ov}$ is mainly
dependent on the distributions of ratios $r_{10}$ and $r_{01}$ that
are only determined by the interior structure \citep{roxb03, oti05}. 
Tables \ref{tab2} $-$ \ref{tab5} list the value of adiabatic oscillation
frequency $\nu_{0}^{m}$ at which $r_{10}$ and $r_{01}$ computed from models 
arrive at the minimum. The values of $\nu_{0}^{m}$ of models with 
$\delta_{\rm ov}$ in the range of $1.7-1.8$ are consistent with that determined
by equation (\ref{o2}). The calculations show that the value of $\nu_{0}^{m}$
is related to the radius of overshooting region of models.
The larger the radius of the core including the overshooting region,
the smaller the value of $\nu_{0}^{m}$.
For KIC 6225718, HD 49933, KIC 2837475, and KIC 11081729
the values of $\nu_{0}$ estimated from observed ratios by using
equation (\ref{o2}) are consistent with those obtained from
$r_{10}$ and $r_{01}$ of the best model of the stars. Thus 
equation (\ref{o2}) aids in understanding the size of the core 
from the observed ratios. If the value of $\nu_{max}$ of a star
is less than the value of $\nu_{0}$ estimated by using 
equation (\ref{o2}), the star may have a relatively small convective
core. However, if the value of $\nu_{max}$ is larger 
than that of $\nu_{0}$, the star may have a large convective core. 

The value of $1.7-1.8$ of the $\delta_{\rm ov}$ is much larger than
the value adopted usually. However, \citet{xiong85} shows that the 
overshooting of convective core could mix material in the distance
of $1.4 H_{p}$. Moreover, the value is consistent with the numerical
result for the downward overshooting of giants \citep{tian09}.
A large $\delta_{\rm ov}$ means that an efficient mixing takes place 
in stellar interior. Rotation can lead to an increase in convective core,
which depends on the efficiency of rotational mixing and rotation 
rate \citep{maed87, yang13a, yang13b}. The effects of rotation explain
the extended MS turnoffs of intermediate-age star clusters in the Large
Magellanic Cloud, where rotational mixing plays an important role 
\citep{yang13b}. The effects of rotational mixing and convection overshoot 
can reconcile the low-Z solar models with helioseismology \citep{yang16}.
Efficient mixing and convection overshoot also may exist in the Sun \citep{yang16}. 
The effect of overshooting can mimic the effects of rotational mixing 
to a certain degree. The surface rotation period of KIC 11081729 is 
about 2.78 days \citep{mcqu14}, which is obviously lower than the 
approximately 27 days of the Sun. The large $\delta_{\rm ov}$ might be related 
to rotation.

\subsection{Summary}
The observed ratios $r_{10}$ and $r_{01}$ of KIC 11081729 decrease firstly
and then increase with frequency. There are different spectroscopic results
for KIC 11081729, which can affect the determinations of mass and age of KIC
11081729 but can not affect the estimation of $\delta_{\rm ov}$. This is due to the fact
that the distributions of $r_{10}$ and $r_{01}$ are mainly dependent on the interior
structure. The structure can be directly changed by overshooting of 
convective core. Thus the distributions of $r_{10}$ and $r_{01}$ are sensitive
to $\delta_{\rm ov}$. For the different spectroscopic constraints, 
the models with $\delta_{\rm ov}$
less than 1.7 can not reproduce the observed ratios. However,
the distributions of the observed ratios can be reproduced well
by models with $\delta_{\rm ov}$ in the range of about $1.7-1.8$.
A large overshooting of convective core may exist in KIC 11081729. 
The behavior of the ratios of KIC 11081729 may derive from the effects 
of the large overshooting of convective core. 

The ratios $r_{10}$ and $r_{01}$ calculated from models with a convective 
core reach a minimum at about $\nu_{0}$, arrive at a maximum at around 
$7\nu_{0}/4$, and then reach a minimum at about $9\nu_{0}/4$. 
These characteristics can be completely reproduced by equation (\ref{o2}).
The distributions of observed and theoretical ratios of stars with 
a convective core can be reproduced well by the equation.
The value of $\nu_{0}$ decreases with the increase in the
radius of convective core and can be estimated by formula (\ref{o2}) 
from observed frequencies. Thus the equation aids in determining the 
presence and the size of convective core including overshooting region
from observed frequencies. If the value of $\nu_{max}$ of a star
is less than the value of $\nu_{0}$ determined from observed frequencies
using equation (\ref{o2}), the star may have a small convective core
and a small $\delta_{\rm ov}$. However, if the value of $\nu_{max}$ is larger 
than that of $\nu_{0}$, the star may have a large overshooting of
convective core.


\begin{ack}
The author acknowledges the support from the NSFC 11273012, 11273007,
11503039, and the HSCC of Beijing Normal University.
\end{ack}


\end{document}